\documentclass[english,twocolumn,showpacs,aps,amsmath,amssymb,preprintnumbers]{revtex4}

\usepackage{graphicx}
\usepackage{dcolumn}
\usepackage{bm}
\usepackage{amscd}
\usepackage{empheq} 

\usepackage[T1]{fontenc}

\usepackage{color}
\usepackage{colordvi}

\usepackage{xspace}
\usepackage{german}
\usepackage{amscd}
\usepackage{xspace}

\newtheorem{corollary}{Corollary}[section]
\newtheorem{lemma}{Lemma}[section]
\newtheorem{proposition}{Proposition}[section]
\newtheorem{remark}{Remark}

\renewcommand{\vec}{\operatorname{vec}}

\newcommand{\R}[1]{\ensuremath{\mathbb R}^{\,#1}{}}
\newcommand{\C}[1]{\ensuremath{\mathbb C}^{\,#1}{}}

\newcommand{\unity}{\ensuremath{{\rm 1 \negthickspace l}{}}}

\newcommand{\RED}[1]{\textcolor{red}{#1}}

\newcommand{\ket}[1]{\ensuremath{| #1 \rangle}{}}

\newcommand{\expt}[1]{\ensuremath{\langle #1 \rangle}{}}
\newcommand{\ad}[1]{\operatorname{ad_{#1}}}

\newcommand{\adr}{\operatorname{ad}}

\newcommand{\Mat}{\operatorname{Mat}{}}
\renewcommand{\vec}{\operatorname{vec}{}}

\newcommand{\uu}{\mathfrak{u}}
\newcommand{\su}{\mathfrak{su}}
\newcommand{\so}{\mathfrak{so}}
\newcommand{\spp}{\mathfrak{sp}}
\newcommand{\sll}{\mathfrak{sl}}
\newcommand{\gl}{\mathfrak{gl}}

\newcommand{\ri}{{\rm i}}

\newcommand{\fe}{\mathfrak{e}}
\newcommand{\ff}{\mathfrak{f}}
\newcommand{\fg}{\mathfrak{g}}
\newcommand{\fk}{\mathfrak{k}}
\newcommand{\fh}{\mathfrak{h}}

\newcommand{\fs}{\mathfrak{s}}
\newcommand{\fu}{\mathfrak{u}}
\newcommand{\fz}{\mathfrak{z}}

\newcommand{\bG}{\mathbf{G}}
\newcommand{\bK}{\mathbf{K}}

 \newcommand{\swap}{\text{\sc swap}\xspace}

\renewcommand{\mod}{\operatorname{mod}{\;}}

\newcommand{\tr}{\operatorname{tr}}

\newcommand{\rank}{\operatorname{rank}{}}
\newcommand{\herm}{\mathfrak{her}{}}

\newcommand{\hoplus}{\ensuremath{\, {\widehat{\oplus}} \,}{}}
\newcommand{\rep}{\ensuremath{\overset{\rm rep}{=}}{}}
\makeatother

\usepackage{babel}

\begin{document}
\preprint{presented in part at the DPG spring conference, Hamburg, March 2.-6. 2009}

\title{Controllability and Observability of Multi-Spin Systems:\\ 
	Constraints by Symmetry and by Relaxation}

\author{U.~Sander and T.~Schulte-Herbr{\"u}ggen}
\email{tosh@ch.tum.de}
\affiliation{Department of Chemistry, Technical University of Munich (TUM), D-85747 Garching,
Germany}

\date{\today}

\begin{abstract}
We investigate the universality of multi-spin systems in architectures
of various symmetries of coupling type and topology. Explicit reachability
sets under symmetry constraints are provided. Thus for a given (possibly symmetric) 
experimental coupling architecture several decision problems can be
solved in a unified way: (i) can a target Hamiltonian be simulated? 
(ii) can a target gate be synthesised? (iii) to which extent is the system 
observable by a given set of detection operators? and, as a special case of
the latter, (iv) can an underlying system Hamiltonian be identified with a
given set of detection operators? Finally, in turn, lack of symmetry
provides a convenient necessary condition for full controllability. 
Though often easier to assess than the well-established Lie-algebra rank condition, 
this is not sufficient unless the {\em candidate} dynamic simple Lie algebra can be pre-identified
uniquely, which is fortunately less complicated than expected.
\end{abstract}
\maketitle

\section*{Introduction}

Controlling the quantum dynamics of experimentally manageable systems 
is paramount to exploiting the great potential quantum systems inherently
promise: they may provide access to 
performing computational tasks or to simulating the behaviour of
other quantum systems that are beyond experimental handling themselves.
Moreover, quantum systems can efficiently simulate classical systems \cite{Fey82, Fey96}
sometimes separating controllable parameters in the quantum analogue
that classically cannot be tuned independently. --- 
Both in simulation and computation
the particular advantage becomes obvious when the complexity of a problem
reduces upon going from a classical to a quantum setting \cite{Kit02}.
On the computational end, most prominently, there is the exponential speed-up by Shor's quantum
algorithm of prime factorisation \cite{Shor94, Shor97} relating 
to the class of quantum algorithms
\cite{Jozsa88, Mosca88} efficiently solving hidden subgroup problems \cite{EHK04}. 
While the demands for accuracy (\/`error-correction threshold\/')
in quantum computation may seem daunting at the moment, the quantum simulation end is by far less sensitive.
Thus simulating quantum systems \cite{Lloyd96}---in particular at phase-transitions \cite{Sachdev99}---has recently
shifted into focus \cite{BCL+02,DNB+02,JC03,PC04} mainly inspired by promising experimental
progress in cold atoms in optical lattice potentials \cite{GMEHB02,BDZ08} as well as in
trapped ions \cite{LBM+03,BW08}. In order to assess feasibility, Kraus {\em et al.}~have explored whether
target quantum systems can be universally simulated on translationally invariant lattices
of bosonic, fermionic, and spin systems \cite{kraus-pra71}. Their work can also be seen as a
follow-up on a study by Schirmer {\em et al.}~\cite{SchiPuSol01} specifically addressing the
question of controllability in systems with degenerate transition frequencies.

Quite generally, quantum control has been recognised
as a key generic tool \cite{DowMil03} needed for advances in experimentally exploiting
quantum systems for simulation or computation (and even more so in future quantum technology). 
It paves the way for constructively optimising strategies 
for experimental implemention in realistic settings. 
Moreover, since such realistic quantum systems are mostly beyond analytical
tractability, numerical methods are often indispensable.
To this end, gradient flows can be implemented on the control amplitudes thus
iterating an initial guess into an optimised pulse scheme \cite{Rabitz87,Krotov,GRAPE}.
This approach has proven useful in spin systems \cite{PRA05} as well as in solid-state 
systems \cite{PRA07}. Moreover, it has recently been generalised from closed systems to 
open ones \cite{PRL_decoh}, where the Markovian setting can also be used as embedding of
explicitly non-Markovian subsystems \cite{PRL_decoh2}.

\medskip

Clearly, for universal quantum computation or simulation, the quantum hardware
(characterised by the {\em system Hamiltonian}) and the 
quantum software (operating via the interface of {\em control Hamiltonians}) have to combine
so that any unitary target operation can be performed irrespective
of the initial state of the quantum system. More precisely, this means the
system has to be fully controllable (as has been pointed out in different 
contexts \cite{RaRa95, Science98}). 
Yet often the quantum hardware comes in designs with a certain symmetry 
pattern reflecting the experimental set up. For instance, this is the case
in quantum lattices, in quantum networks, or in spin chains serving as 
quantum wires \cite{Bose07} for distributed quantum computation \cite{vMNM07}.
However, symmetric patterns may have crucial shortcomings since symmetry may 
preclude full controllability. On the other hand, avoiding 
symmetry-restricted controllability need not be complicated from an experimental 
point of view: in Ising spin chains it soon emerged that polymers $(ABC)_n$ made
of three different qubit units $A,B,C$ are fully controllable \cite{Lloyd93, Lloyd99}. 
Later irregular $ABAAA\dots$ systems of just two qubit types $A,B$ \cite{FT06} 
and even $A-A\cdots A-B$ systems 
turned out to be fully controllable as well. In these systems, spin qubits
are meant to be locally controlled by operations that act jointly on all
qubits labelled with the same letter and independently from controls on qubits
with a different letter.
More specifically, quantum systems coupled by Heisenberg $XXX$ type interactions turned
out to be fully controllable when the local controls are limited to a small subset of
qubits, as with time these actions can then be \/`swapped\/' to neighbouring spins
\cite{B00,Bose03,CDE+04,EPB+04,Burg05,Bose07,SPP08}. 

Given the power of Lie theory to assess controllability as well as observability
in classical systems \cite{SJ72, JS72},
such a gradual case-by-case development asks for a more systematic investigation on the quantum
side.  Here we address controllability and observability with and without symmetry restrictions 
in multi-qubit systems, where the
coupling topology is generalised to any connected graph \cite{TOSH-Diss, AlbAll02, Burg08}.
In doing so we also specify reachability sets as subgroup orbits 
for systems that are not fully controllable since it is important to know for
which dedicated tasks of simulation or computation symmetry restricted systems can still be used.

{\em Scope and Overview:}\quad 
On a more general scale, it is important to be able to separate questions of 
existence (e.g.: {\em is the system fully controllable? is it observable? can it be used for universal 
Hamiltonian simulation?}) 
from questions of actual implementation ({\em how does one have to steer a given
experimental setup for implementing a target task with highest precision?}). 
Otherwise the elegance of constructive proofs of existence 
may all too often come at the cost of highly \/`suboptimal\/' implementation when
translated directly into experimental schemes irrespective of the actual given setting.
Therefore, here we advocate to exploit the power of Lie theory for a unified framework 
addressing controllability (and within the context of tomography also observability) 
in a first step, while resorting to
quantum control in a second step for actual implementation optimised for the
given experimental set up. 

In view of practical applications, we also systematically explore 
what we call {\em task controllability}, i.e. which tasks are feasible 
on which type of quantum system hardware. In turn, the unified picture also
provides the rules how to design quantum hardware for a specific task.

\section{Controllability}
Here we address Markovian dynamics of quantum systems, the free evolution of which 
is governed by a {\em system Hamiltonian} $H_d$ and, in the 
case of open systems, by an additional {\em Markovian relaxation term} $\Gamma$ taking 
GKS-Lindblad form. Whenever we talk about controllability, we mean {\em full operator
controllability} thus neglecting more specialised notions like pure-state controllability
\cite{AA03}.

\subsection{Quantum Dynamical Control Systems}

The interplay between the quantum system and the experimenter
is included by {\em control Hamiltonians} $H_j$ expressing external manipulations in 
terms of the quantum system itself, where each control Hamiltonian can be
steered in time by {\em control amplitudes} $u_j(t)$.
With these stipulations the usual equations of motion for controlled quantum dynamics 
can be brought into a common form as will be shown next. 

To this end, consider the Schr{\"o}dinger equations
\begin{eqnarray}
        \ket{\dot\psi(t)} &=& -i\big(H_d + \sum_{j=1}^m u_j(t) H_j\big) \;\ket{\psi(t)}\\
        \label{eqn:bilinear_contr}
        {\dot U(t)} &=& -i\big(H_d + \sum_{j=1}^m u_j(t) H_j\big) \;{U(t)} \quad,
\end{eqnarray}
where the second identity can be envisaged as the operator equation to the first
one. It governs the evolution of a unitary map of an entire basis set
of vectors representing pure states. Using the short-hand notations
$H_u:= H_d + \sum_{j=1}^m u_j(t) H_j$ and $\adr_H(\cdot):=[H,(\cdot)]$
also consider the master equations 
\begin{eqnarray}\label{eqn:master}
\dot{\rho}(t) &=& -i [H_u,\rho(t)] - \Gamma(\rho(t)) \equiv -(i \adr_{H_u} \,+\,\Gamma)\; \rho(t)\quad\\[2mm]
\dot F(t)  &=& -(i \adr_{H_u} \,+\,\Gamma)\;\circ\; F(t) \quad.\label{eqn:super_master}
\end{eqnarray}
While $\rho\in \herm(N)$, $F$ denotes a {\em quantum map} in $GL(N^2)$ as linear image over all basis
states of the Liouville space representing the open system.

All these equations of motion ressemble the form of a standard 
{\em bilinear control system} $(\Sigma)$ known in classical systems and control theory
reading
\begin{equation}
        \dot X(t) = \big(A + \sum_{j=1}^m u_j(t) B_j\big) \; X(t) 
\end{equation}
with \/`state\/' $X(t)\in\C N$, drift $A\in \Mat_N(\C{})$, controls $B_j\in \Mat_N(\C{})$, 
and control amplitudes $u_j\in\R{}$.
For simplicity consider for the moment its {\em linear} counterpart
with $v\in\C N$
\begin{equation}
        \dot X(t) = A X(t) + B v
\end{equation}
which is known to be {\em fully controllable} if it obeys 
the celebrated rank condition (see, e.g., \cite{LeeMarkus})
\begin{equation}
	\rank [B, AB, A^2B, \dots, A^{N-1}B] = N\quad.
\end{equation}
Now lifting the bilinear control system $(\Sigma)$ to group manifolds \cite{Bro72,Jurdjevic97} 
by $X(t) \in GL(N,\C{})$ under the action of some compact connected Lie group $\mathbf K$ with Lie algebra $\fk$ 
(while keeping $A,B_j\in \Mat_N(\C{})$),
the condition for full controllability turns into its analogue known as the
{\em Lie algebra rank condition} \cite{SJ72, JS72, Jurdjevic97}
\begin{equation}
        \langle A, B_j \,|\,j=1,2,\dots,m\rangle_{\rm Lie} = \fk\quad,
\end{equation}
where $\langle \cdot \rangle_{\rm Lie}$ denotes the {\em Lie closure} obtained by
repeatedly taking mutual commutator brackets.

\subsection{Closed Quantum Systems: Full Controllability and Symmetry-Restricted Controllability}
Clearly in the dynamics of closed quantum systems, the system Hamiltonian $H_d$ is the only 
drift term, whereas the $H_j$ are again the control Hamiltonians.
To fix notations, in systems of $n$ qubits define $N:=2^n$, so these traceless 
spin Hamiltonians $i\,H_\nu\in\mathfrak{su}(N)$
each generate a one-parameter unitary group of time evolution $U_\nu(t):=e^{-itH_\nu}\in SU(N)$. 

Transferring the classical result \cite{JS72} to the quantum domain \cite{RSD+95}, 
the bilinear system of Eqn.~\ref{eqn:bilinear_contr} is {\em fully operator controllable} 
if and only if the drift and controls are a generating set of $\mathfrak{su}(N)$ 
\begin{equation}
\langle{i H_d, i H_j} \,|\,j=1,2,\dots,m\rangle_{\rm Lie} = \fk = \mathfrak{su}(N)\quad.
\end{equation}
In fully controllable systems to every initial state $\rho_0$ the {\em reachable set} is 
the entire unitary orbit $\mathcal O_{\rm U}(\rho_0):=\{U\rho_0 U^\dagger\;|\; U\in SU(N)\}$.
With density operators being Hermitian
this means any final state $\rho(t)$ can be reached from any initial state $\rho_0$
as long as both of them share the same spectrum of eigenvalues.

In contrast, in systems with restricted controllability the Hamiltonians
only generate a proper subalgebra 
\begin{equation}
\langle{i H_d, i H_j} \,|\,j=1,2,\dots,m\rangle_{\rm Lie} = \fk \subsetneq \mathfrak{su}(N)\quad.
\end{equation}

\subsubsection*{Algorithm for Computing the Lie Closure}
\begin{figure}[Ht!]
\begin{tabular}{llllll}
\hline\hline\\[-1mm]
\multicolumn{6}{l}{{\bf Algorithm 1} \cite{SchiFuSol01}: Determine Lie closure for $n$-qubit system}\\
\multicolumn{6}{l}{with given set of drift (or system) and control Hamiltonians}\\[1mm]
\hline\\[-1mm]
&&&1. Start with the inital basis $B_\nu:=\{H_d; H_1,\dots, H_m\}$.&&\\[2mm]
&&&2. If $m+1=\dim\, \su(2^{n})$ $\Rightarrow$ terminate.&&\\[2mm]
&&&3. Perform all Lie brackets $K_{i}=[H_{j},H_{k}]$.&&\\[2mm]
&&&4. For each new $K_i$ check linear independence&&\\
&&&   \phantom{4. }from ${\rm span}\, B_\nu$. If nothing new found $\Rightarrow$ terminate.&\\[2mm]
&&&5. Extend basis by all independent new $\{K_{i}\}$&&\\
&&&\phantom{5. }$B_{\nu+1}:=\{\{K_{i}\},H_{i}\}$. Go to (1).&&\\[2mm]
\hline\hline
\end{tabular}
\end{figure}

Given an $n$-qubit bilinear control system characterised by the 
drift and control Hamiltonians $\{H_d; H_1,\dots, H_m\}$.
Then the canonical {\bf Algorithm 1} tabulated above
constructively determines a basis of the associated
dynamic Lie algebra $\fk$ \cite{SchiFuSol01}. --- Our implementation codes the
orthogonal basis of tensor products of Pauli matrices as quaternions simply 
represented by the Clifford algebra $\mathcal C\ell_2(\mathbb R)$ 
of quarternary numbers $\{0,1,2,3\}$
plus the Clifford multiplication rule allowing to calculate Lie brackets
without matrix representations. The only time consuming step is
the final rank determination by QR decomposition of a sparse coefficient
matrix $C\in\Mat_{4^n}$ collecting all the expansion coefficients to the 
$K_i$ of {\bf Algorithm 1} columnwise as $\vec(K_i)$ \cite{HJ2}. Our
results have been cross-checked with {\sf{GAP 4.4.12}} \cite{GAP4}.

%
%
%
%
%
%
%
%

\subsubsection*{Notation: Coupling Graphs}
Here we adopt the notation 
to represent a system by a graph $G(V,E)$ whose vertices represent the qubits and
whose edges denote pairwise couplings of Ising or Heisenberg nature.
In the subsequent examples, 
the qubits are taken to be locally controlled by operations that act jointly on all
qubits labelled with the same letter and independently from controls on qubits
with a different letter. We assume that every qubit type is fully controllable
so all $SU(2)$-actions are jointly admissible on each qubit with the same letter.
Let us emphasise that although the current set of examples is taken with some local control on 
each qubit, the algebraic approach put forward here extends to systems with subsets of 
uncontrolled qubits coupled by isotropic
Heisenberg interactions thus reproducing all the results on transmitting controls
through infective graphs given recently \cite{Burg08}.

\subsubsection*{Characterisation by Symmetry}
\begin{figure}[Ht!]
\begin{centering}
\includegraphics[scale=0.5]{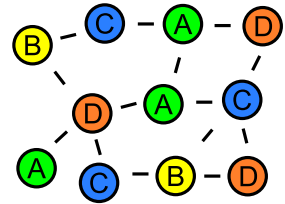}
\par\end{centering}
\caption{General coupling topology represented by a connected graph. The vertices denote the spin-$\tfrac{1}{2}$ 
qubits, while the edges represent pairwise couplings of Heisenberg or Ising type.
Qubits of the same colour and letter are taken to be controlled by joint local
unitary operations as in Tab.~I (or none: see Tab.~II), while qubits of different kind can be
controlled independently. For a system to show a certain symmetry (brought about by
permuting subsets of qubits), it is necessary
that both the graph as well as the system plus all control Hamiltonians are invariant,
see text.\label{fig:gen-topo-graph}}
\end{figure}

In the following, we will characterise systems of restricted controllability in terms of 
symmetries. In the present setting, a Hamiltonian quantum system is said to have 
a {\em symmetry} expressed by the skew-Hermitian operator $s \in \mathfrak{su}(N)$, if
\begin{equation}
[s, H_\nu] = 0 \quad\text{for all}\quad \nu \in \{d; 1,2,\dots,m\}\quad.
\end{equation}
More precisely, we use the term {\em outer symmetry} if $s$ 
generates a SWAP operation permuting a subset of spin qubits such that the coupling 
graph is left invariant.
In contrast, an {\em inner symmetry} relates to elements $s$ 
not generating a SWAP operation in the symmetric group of all permutations of
spin qubits in the system.

In either case, a symmetry operator is an element of the centraliser (or synonymously the commutant)
\begin{equation*}
\{H_\nu\}':= \big\{s\in\mathfrak{su}(N)\,|\, [s, H_\nu] = 0 \; \forall \nu \in \{d; 1,2,\dots,m\}\big\}\,. 
\end{equation*}
Recall that the {\em centraliser} or {\em commutant} of a 
fixed subset $\mathfrak m\subseteq \mathfrak g$ with respect to a Lie algebra $\mathfrak g$
consists of all elements in $\mathfrak g$ that commute with all elements in $\mathfrak m$.
By Jacobi's identity \mbox{$\big[[a,b],c\big]+\big[[b,c],a\big]+\big[[c,a],b\big]=0$}
one gets two properties of the commutant pertinent for our context: 
First, an element $s$ that commutes with the Hamiltonians $\{iH_\nu\}$ also commutes 
with their Lie closure $\fk$. For the dynamic Lie algebra $\fk$ we have
\begin{equation}
\fk' := \{s \in \mathfrak{su}(N)\;  |\; [s,k] = 0 \;\;\forall k \in \fk\} 
\end{equation}
and $\{iH_\nu\}'= \fk'$.
Thus in practice it is (most) convenient to just evaluate the commutant 
for a (minimal) generating set $\{iH_\nu\}$
of $\fk$. Second, for a fixed $k\in\fk$ an analogous argument gives
\begin{equation*}
[s_1,k] = 0 \quad \text{and}\quad [s_2,k] = 0 \quad \Longrightarrow\quad \big[[s_1,s_2],k\big]\quad,
\end{equation*}
so the commutant $\fk'$ forms itself a Lie subalgebra to $\mathfrak{su}(N)$ 
consisting of all symmetry operators.
Since the commutant $\fk'$ to the dynamic Lie algebra $\fk$ is invariant under $\fk$,
it is a {\em normal subalgebra} or an {\em ideal} to $\fk$
\begin{equation}\label{eqn:comm-id}
[\fk, \fk'] \subseteq \fk' \quad.
\end{equation}
NB: while the embedding Hilbert space $\mathcal H$ of Hermitian operators is endowed with a
weak (and strong) operator topology thus giving von Neumann's bicommutant theorem \cite{vNeu29} 
$\mathcal H'' = \mathcal H$ well-known in $C^*$-algebras \cite{Dix77,Sak71}, no such
theorem extends to Lie algebras: there may be 
several inequivalent Lie algebras sharing the same commutant, in particular when the 
commutant is trivial.

Within the commutant $\fk'$ one may choose its centre 
$\fz = \fk'\cap\fk''$
of mutually commuting symmetry operators
then allowing for a block-diagonal representation in
the eigenspaces associated to
the eigenvalues $(\lambda_1, \lambda_2, \dots, \lambda_\ell)$ to
$\{z_1, z_2, \dots , z_\ell\}=\fz$.
A convenient set of symmetry operators embracing the outer symmetries
are the ones generating SWAP
transpositions of qubits: they correspond to the $S_2$ symmetry and
come with the eigenvalues $+1$ ({\em gerade}) and $-1$ ({\em ungerade}).
A block-diagonal representation results if all the SWAP transpositions
that can be performed independently are taken as one entry each in the $\ell$-tuple
$(\lambda_1, \lambda_2, \dots, \lambda_\ell)$, while all those that have to
be performed jointly are multiplied together to make one single entry in the tuple.
This procedure is illustrated below in Examples 1 and 2.

\subsubsection*{Notation: Clebsch-Gordan Decomposition}
Observe that in our notation a {\em block-diagonal representation} of a Lie algebra
$\fk = \su(N_1)\oplus \su(N_2)$ generates a group 
$\bK = SU(N_1)\oplus SU(N_2)$ in the sense of a block-diagonal Clebsch-Gordan decomposition, 
while 
the abstract {\em direct sum} of the algebras has a matrix representation as the {\em Kronecker sum}
$\fg=\su(N_1) \hoplus \su(N_2):=\su(N_1)\otimes\unity_{N_2} + \unity_{N_1}\otimes\su(N_2)$ 
and generates a group isomorphic to the {\em tensor product} $\bG=SU(N_1)\otimes SU(N_2)$. 
Note that the direct sum of two algebras $\fg$ and $\fh$ (each given in an irreducible
representation) has itself an irreducible representation as Kronecker 
sum $\fg \hoplus \fh$ \cite{Cornwell:1984}.

\bigskip
\subsection*{Examples with Symmetry-Restricted Controllability}
\subsubsection*{Example 1: Joint $S_{2}$ Symmetry}
\begin{figure}[Ht!]
\begin{centering}
\includegraphics[scale=0.5]{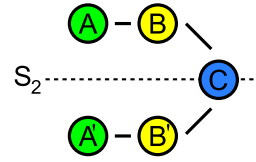}
\par\end{centering}
\caption{Ising spin chain with joint $S_2$ symmetry.\label{fig:scheme-abcba-joint}}
\end{figure}
\begin{figure}[Ht!]
\begin{centering}
\includegraphics[scale=0.2]{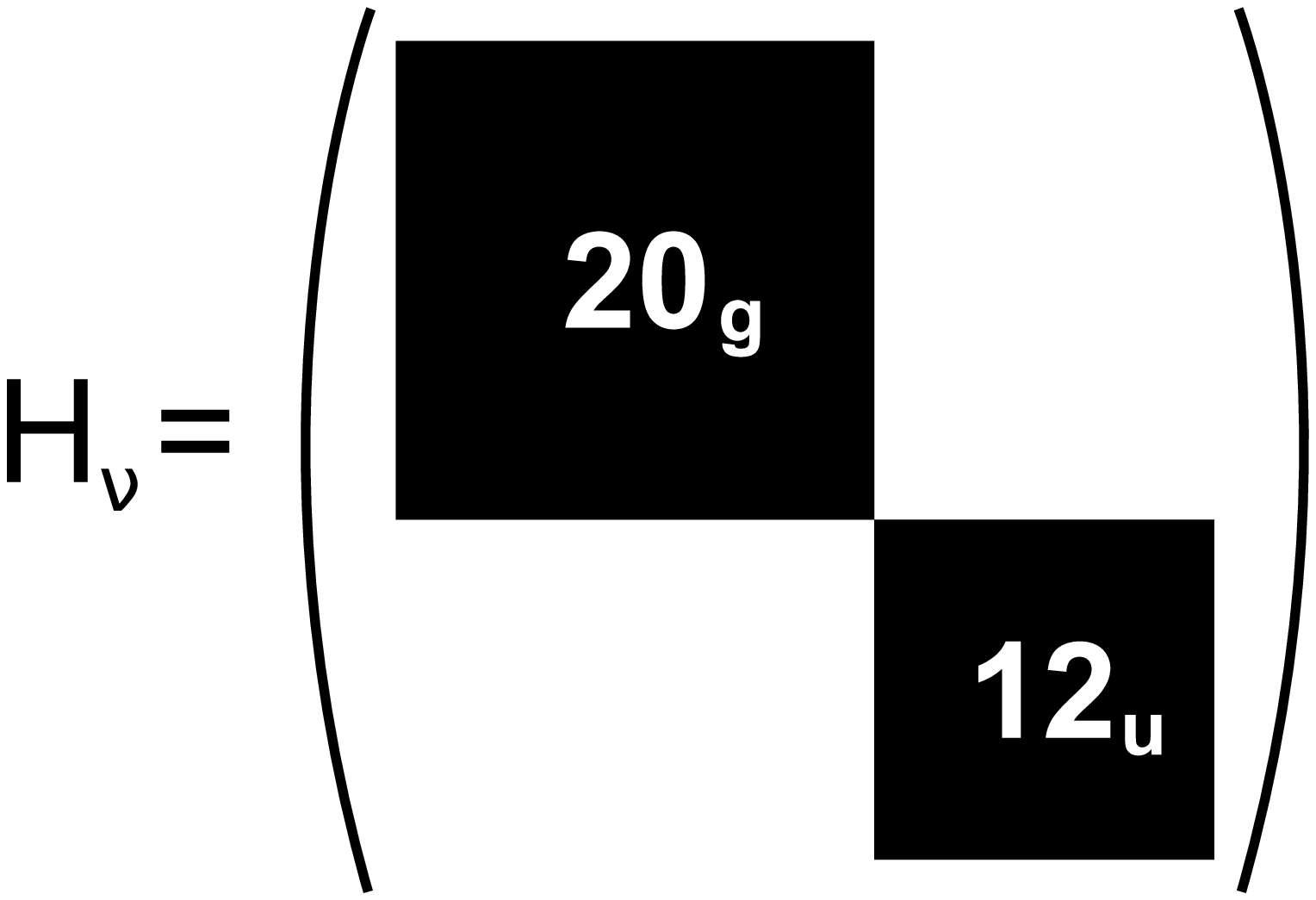}
\par\end{centering}
\caption{The drift and control Hamiltonians of Ex.~1 take block diagonal form 
corresponding to the $A_g$ and $A_u$ representation of the 
$S_2$-symmetry group.\label{fig:block-abcba-global2}}
\end{figure}

First, consider Ising $n$-spin-$\tfrac{1}{2}$ chains with odd numbers of spins such as, e.g., the one
in Fig.~\ref{fig:scheme-abcba-joint}, which has an Ising coupling graph $L_5$. 
Denote the centre spin by $C$.
An $S_{2}$-symmetry leaves the coupling graph and thus all Hamiltonians
of drifts and controls $\{H_d,H_1,\dots,H_m\}$ invariant under the \emph{joint} permutation 
$j\leftrightarrow (N-j+1)$ \quad$\forall\; j=1\dots\lfloor n/2\rfloor$.
In the $S_2$-symmetry-adapted basis, the Hamiltonians take block-diagonal form
with two blocks of different parity, \textbf{\emph{g}}\emph{erade}
and \textbf{\emph{u}}\emph{ngerade}. 
The dimensions of the respective blocks are
\begin{eqnarray}
d_{g} & :=& 2^{n}\sum_{k=a}^{(m+a)/2}\left(\begin{array}{c} m\\
2k-a\end{array}\right)3^{m+a-2k}\\
d_{u} & :=& 2^{n}-d_{g}
\end{eqnarray}
where $m$ denotes the number of symmetric spin pairs and $a:=m(\mod 2)$.
Moreover, each block represents a fully controllable logical subsystem.

In the system depicted in Fig.~\ref{fig:scheme-abcba-joint},
spins $A$ and $B$ can be jointly exchanged with spins $A'$ and $B'$, 
since the system has a mirror symmetry with $C$ in the mirror axis.
As shown in Fig.~\ref{fig:block-abcba-global2},
the $S_2$-symmetry adapted basis entails block diagonal representations of
the Hamiltonians with one block being of size $20\times20$
and one block of size $12\times12$. Note that all the symmetrised 
Hamiltonians are traceless within each of the two blocks.
The Lie closure gives a Clebsch-Gordan (CG) decomposed dynamic Lie algebra
\begin{equation}
\fk_1 = \su(20)\oplus\su(12) \quad\text{so}\quad \bK_1 = SU(20)\oplus SU(12)
\end{equation}
and thus the Lie rank is $542 = 399 + 143$.

\subsubsection*{Example 2: Individual Permutation Symmetry}

\begin{figure}[Ht!]
\begin{centering}
\includegraphics[scale=0.5]{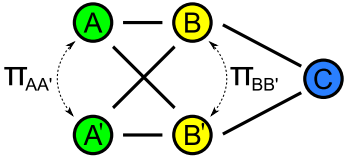}
\par\end{centering}
\caption{Coupling topology allowing for individually independent 
permutation symmetries $\Pi_{AA'}$ and $\Pi_{BB'}$, which
together reproduce the $S_2$-symmetry of Ex.~1.
\label{fig:scheme-abcba-pairwise}}
\end{figure}
\begin{figure}[Ht!]
\begin{centering}
\includegraphics[scale=0.2]{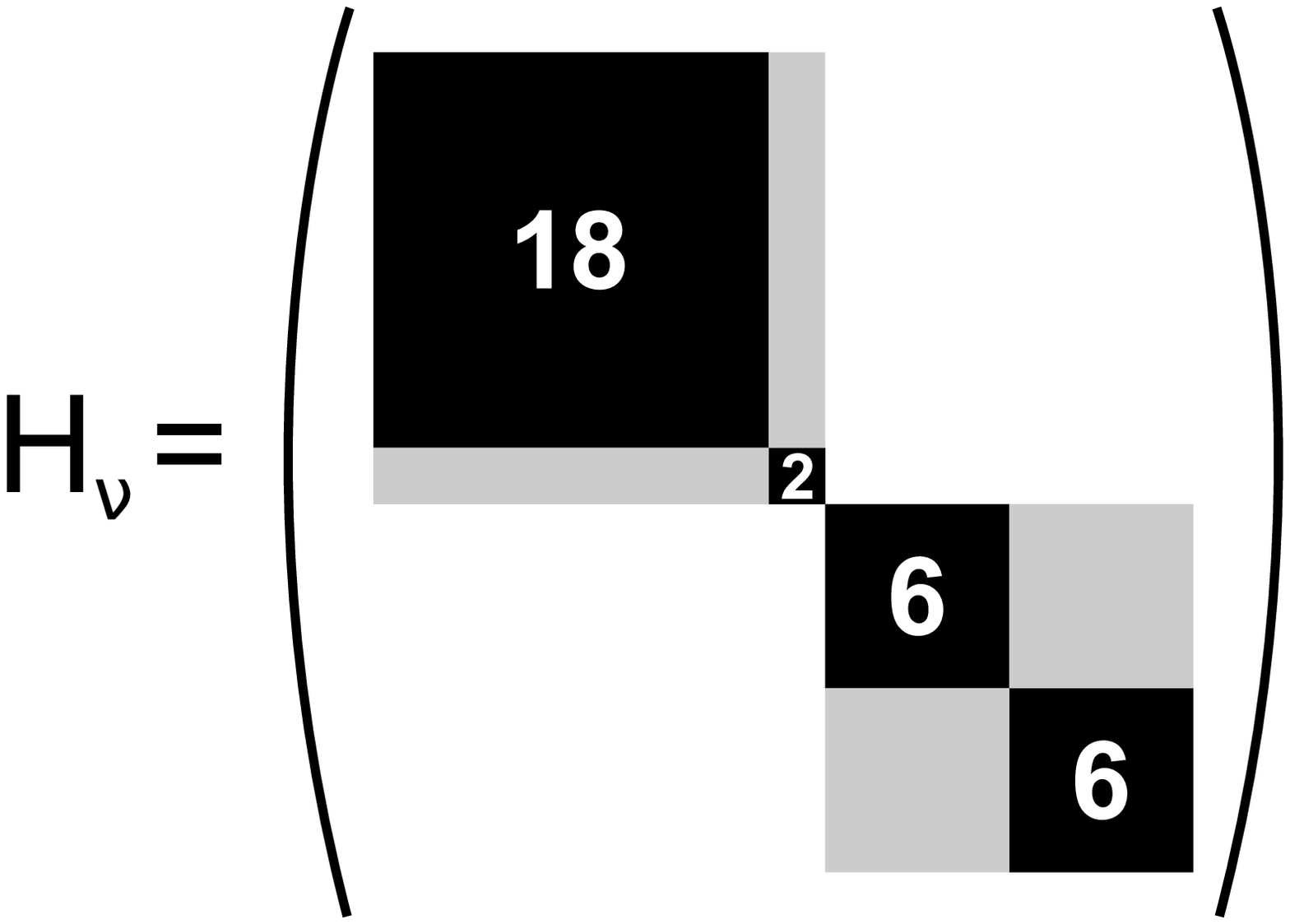}
\par\end{centering}
\caption{The drift and control Hamiltonians of Ex.~2 take block diagonal form
corresponding to the $gg$, $uu$, $gu$ and $ug$ parities of the individual 
permutations $\Pi_{AA'}$ and $\Pi_{BB'}$.\label{fig:blocks-abcba-fine}}
\end{figure}
\begin{table*}[Ht!]
\caption{\label{tab:ex} Dynamic System Algebras to Qubit Systems with Joint Local Controls 
	on Each Type (Letter) and Various Couplings}
\begin{tabular}{lcccccc}
\hline\hline\\[-1mm]
System & Coupling Types & Lie Rank & Block Sizes & Lie Ranks & System Lie Algebra & Trace = 0 \\
       &                &          &             & blockwise &                    & blockwise \\[1mm]
\hline\\[-1mm]
$A-A$ & $ZZ, XX, XY$ & 9 & 3, 1 & 9, 1 & $\fs(\fu(3)\oplus \fu(1))$ & no \\[2mm]
$A-A$ & $XXX$ & 4 & 3, 1 & 4,1 & $\fs(\fu(2)\oplus \fu(1))$ & no \\[2mm]
$A-B$ & --all-- & 15 & 4 & 15 & $\su(4)$ & yes \\[2mm]
$A\mid B$ & --none-- & 6 &  4 & 6 & $\su(2)\hoplus \su(2)$ & yes \\[2mm]\hline\\[-1mm]
$A-B-A$ & --all-- & 38 & 6, 2 & 35, 3 & $\su(6)\oplus\su(2)$ & yes \\[2mm]\hline\\[-1mm]
$A-B-B-A$ & --all-- & 135 & 10, 6 & 100, 36 & $\fs(\fu(10) \oplus \fu(6))$ & no \\[2mm]
$\underbracket[.4pt]{A-B-B-A}$ & $ZZ, XX, XY$ & 135 & 10, 6 & 100, 36 & $\fs(\fu(10)
\oplus \fu(6))$ & no \\[2mm]
$\underbracket[.4pt]{A-B-B-A}$ & $XXX$ & 115 & 10, 6 & 100, 16 & $\fs(\fu(10) \oplus \fu(4))$ & no \\[2mm]
\hline\\[-1mm]
$A-B-C-B-A$ & --all-- & 542 & 20, 12 & 399, 143 & $\su(20)\oplus\su(12)$ & yes \\[2mm]
$\underbracket[.4pt]{A-B-C-B-A}$ & --all-- & 543 & 20, 12 & 400, 144 & $\fs(\fu(20)
\oplus \fu(12))$ & no \\[2mm]
$A - \underbracket[.4pt]{B - C - B
\makebox[0pt][r]{$\displaystyle{\overbracket[.4pt]{\phantom{A-B-C-B}}}$}
- A}$ & --all-- & 364 & 18, 2, 6, 6 & \quad 323, (3), 35, 6 & %
	\quad given in lengthy Eqn.\eqref{eqn:k2}\quad\phantom{.} & yes \\[2mm]
$\underbracket[.4pt]{A-B-C-B}-A$ & --all-- & 1023 & 32 & 1023 & $\su(32)$ & yes \\[2mm]\hline\\
$A-B-C-C-B-A$ & all & 2079 & 36, 28 & 1296, 784 & $\fs(\fu(36)\oplus\fu(28))$ &
no \\[2mm]
\hline\hline
\end{tabular}
\end{table*}

\begin{figure}[Ht!]
\begin{centering}
\includegraphics[scale=.6]{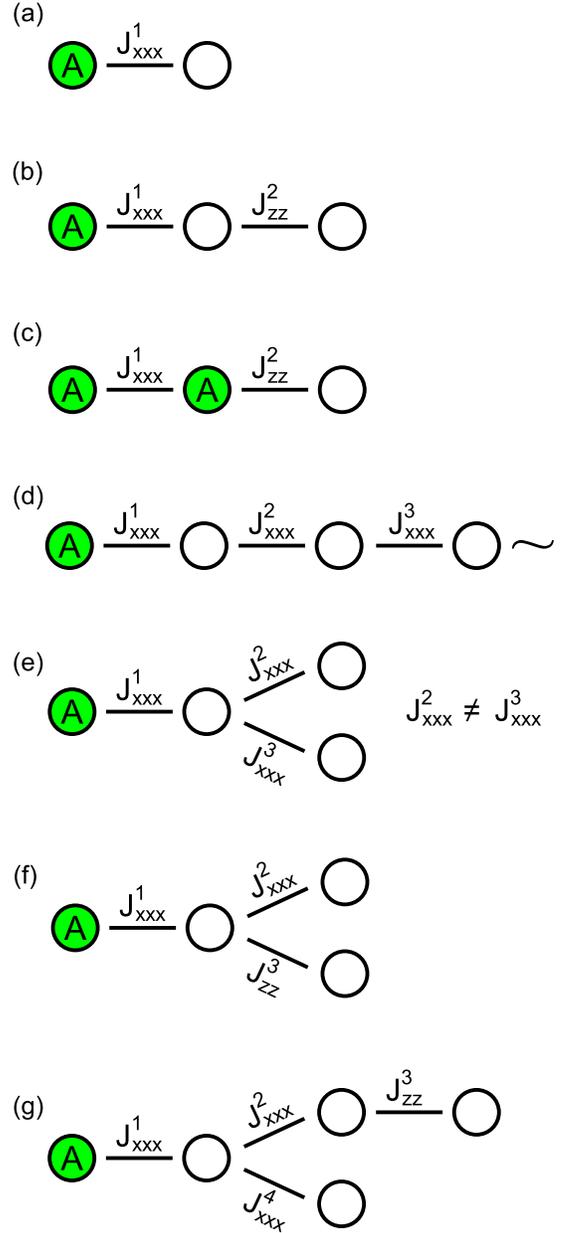}
\par\end{centering}
\caption{Coupled qubit systems with subsets of uncontrolled spins that are coupled via a connected coupling topology
of Heisenberg interactions. Their dynamic Lie algebras are listed in Tab.~II. Note that the graph of 
(e) is \/`non infective\/' \cite{Burg08}, yet the system is fully controllable. \label{fig:tab2fig}}
\end{figure}

\begin{table*}[Ht!]
\caption{\label{tab:ex2} Dynamic System Algebras to Qubit Systems with Proper Subsets of Uncontrolled Qubits}
\begin{tabular}{lcccccc}
\hline\hline\\[-1mm]
System & \quad Coupling Types\quad\phantom{.}  & \quad Lie Rank\quad\phantom{.}  & \quad Block Sizes\quad\phantom{.} & \quad Lie Ranks\quad\phantom{.}  & System Lie Algebra\qquad\phantom{.}  & \quad Trace = 0\quad\phantom{.}  \\
shown in &  non-zero $J_i$ &          &             &  blockwise &                    & blockwise \\[1mm]
\hline\\[-1mm]
Fig.~\ref{fig:tab2fig}(a) & $J^1_{\rm arb.}$ & 15 & 4 & 15 & $\su(4)$ & yes \\[2mm]
\hline\\[-1mm]
Fig.~\ref{fig:tab2fig}(b) & $J^1_{xxx}, J^2_{zz}$ & 30 & 4, 4 & 15, 15 & $\su(4)\oplus \su(4)$ & yes \\[2mm]
Fig.~\ref{fig:tab2fig}(c) & $J^1_{xxx}, J^2_{zz}$ & 30 & 4, 4 & 15, 15 & $\su(4)\oplus \su(4)$ & yes \\[2mm]
\hline\\[-1mm]
Fig.~\ref{fig:tab2fig}(d) & $J^k_{xxx}$ & 255 & 16 & 255 & $\su(16)$ & yes \\[2mm]
Fig.~\ref{fig:tab2fig}(e) & $J^1_{xxx}, J^2_{xxx}\neq J^3_{xxx}$ & 255 & 16 & 255 & $\su(16)$ & yes \\[2mm]
Fig.~\ref{fig:tab2fig}(f) & $J^{1,2}_{xxx}, J^3_{zz}$ & 126 & 8, 8 & 63, 63 & $\su(8)\oplus\su(8)$ & yes \\[2mm]
\hline\\[-1mm]
Fig.~\ref{fig:tab2fig}(g) & $J^{1,2,3}_{xxx}, J^4_{zz}$ & 510 & 16, 16 & 255, 255 & $\su(16)\oplus\su(16)$ & yes \\[2mm]
\hline\hline
\end{tabular}
\end{table*}

By introducing further Ising couplings between spins $A$ and $B'$
and spins $A'$ and $B$, the $L_5$ system of Ex.~1
can be turned into the one represented by a non-planar graph 
in Fig.~\ref{fig:scheme-abcba-pairwise}:
note that now the spin pairs $A,A'$ and $B,B'$ can be permuted individually.
Thus the block-diagonal representation consists of four blocks
corresponding to the parities $gg$, $uu$ and $gu$, $ug$ with
the respective sizes 
$18\times18$, $2\times2$ and twice
$6\times6$, see Fig. \ref{fig:blocks-abcba-fine}.
The indistinguishable pair $AA'$ (and $BB'$) can be looked upon as a
pseudo spin-1 and pseudo spin-0 system, because symmetrisation allows for
adding their spin angular momenta in the usual Clebsch-Gordan way.
The Lie rank of each pair is 4 instead of 15 in
a fully controllable spin pair. 

Now the Lie rank of the total system is $364=323+35+6$, i.e. the sum
of all blockwise Lie ranks---with two exceptions:
(i) the second $6\times6$ block
is not of full rank since it does not contain any coupling terms;
instead, it only collects elements from two {\em independent} $\su(2)$
subalgebras which arise from the $AA'$ pair and the $C$ spin
thus giving a Lie rank of $6$;
(ii) the $2\times2$ block does not contribute since it reduplicates
one of the independent $\su(2)$ algebras 
already occuring in the second $6\times6$ block, which becomes obvious as the
matrix elements in both blocks only occur jointly.
Finally, since in all Hamiltonian components all the blocks are independently traceless,
we have a CG-decomposed dynamic algebra 
\begin{equation}\label{eqn:k2}
\fk_2 = \su(18)\oplus\su\underbracket[.4pt]{(2)\oplus\su(6)\oplus\big({\su(2)}_{j=1}\hoplus\su}(2)\big)
\end{equation}
generating a dynamic group 
\begin{equation}
\bK_2  =SU(18)\oplus SU\underbracket[.4pt]{(2) \oplus SU(6)\oplus\big({SU(2)}_{j=1}\otimes SU}(2)\big)\quad,
\end{equation}
where the index $j$=1 denotes the spin-1 representation of $\su(2)\subset\su(3)$ and the bracket
connects the $\su(2)$ of spin $C$ occuring in two copies ({\em vide supra}).


\subsection{Task Controllability}
Clearly a set of non-trivial symmetry operators precludes full controllability
since the symmetry restrictions entail that the dynamic Lie algebra
$\fk$ is but a proper subalgebra of $\su(N)$. Thus
\begin{equation}
\langle\exp\fk\rangle=:\mathbf K \subsetneq SU(N)
\end{equation}
and the reachable sets take the form of {\em subgroup orbits}
\begin{equation}
\mathcal{O}_{K}(\rho_0):=\{K\rho_{0}K^{-1}|\, K\in \mathbf K\}\quad.
\end{equation}
This is one reason why in the case of symmetry-reduced controllability
it is advantageous to have specified the dynamic Lie
algebra $\fk$ explicitly by the methods introduced above.
Another reason is to be able to assess the feasibility of specific tasks
in the case of non universal systems of reduced controllability. 
More precisely, a Hamiltonian quantum system characterised by $\{iH_\nu\}$ is
called to be {\em task controllable} with respect to a target unitary module
(or gate) $U_T$ if there is at least one Hamiltonian $H_T$ on some branch 
of the \/`logarithm\/' of $U_T$ so that $U_T=e^{i\phi}\cdot e^{-i H_T}$ 
(with arbitrary phase $\phi$) and $i H_T \in \fk = \langle iH_\nu \rangle_{\rm Lie}$.
For a detailed discussion of the phase and its experimental 
significance in systems with time hierarchies between local and coupling
operations see Ref.~\cite{PRA05}. Needless to say only fully controllable systems are
task controllable with respect to all unitary target modules.

\subsection{Lack of Symmetry {\em versus} Full Controllability}
Ultimately the question arises under which conditions the {\em absence of any symmetry} in turn 
implies full controllability. In the special case of pure-state controllability,
this interrelation was analysed in \cite{AlbAll02}.
In the generalised context of full operator controllability 
the issue has been raised in \cite{dAless08}, {\em inter alia}
following the lines of \cite{TR03}, however, without a full answer.
In particular, here we ask: what is special about quantum systems where the 
drift Hamiltonian comprises Ising or Heisenberg-type couplings in a topology that can take 
the form of any {\em connected} graph? ---
To this end, observe the following:

\begin{lemma}
Let $\fk\subseteq\su(N)$ be a matrix Lie subalgebra to the compact simple Lie algebra of
special unitaries $\su(N)$. If its commutant (centraliser) $\fk'$ of $\fk$ in $\su(N)$
is trivial, then 
\begin{enumerate}
\item[(1)] $\fk$ is irreducible;
\item[(2)] $\fk$ is simple or semi-simple.
\end{enumerate}
\end{lemma}
{\bf Proof.}
(1)
The unitary representation as matrix Lie algebra $\fk\subseteq\su(N)$ is 
fully reducible by the Schur-Weyl theorem. Now as there is no invariant subspace $\fk'$ other than
the trivial ones, the representation is irreducible.\\
(2) Since $\fk$ is by construction a Lie subalgebra to the compact Lie algebra $\su(N)$ it is
compact itself. By compactness it has a decomposition into its centre and a semi-simple part 
$\fk = \fz_\fk \oplus \fs$ (see, e.g.,~\cite{Knapp02} Corollary IV.4.25). As the centre
$\fz_\fk = \fk' \cap \fk$ is trivial, $\fk$ can only be semi-simple or simple.
\hfill$\blacksquare$

\medskip
Moreover, by considering invariant subalgebras (ideals) one gets further
structural properties.


\begin{lemma}
Let the Lie closure $\fk$ of a set of drift and control Hamiltonians $\{i H_\nu\}$ 
be a compact Lie algebra with trivial commutant $\fk'$ in $\su(N)$. Then
$\fk$ is a simple Lie algebra, if there is a drift Hamiltonian $i H_d\in \{i H_\nu\}$
satisfying the two conditions
\begin{enumerate} 
\item the coupling topology to $i H_d$ takes the form of a {\em connected} graph
	extending over all the qubits; 
\item the coupling type is Ising $ZZ$ or Heisenberg 
$XX$, $XY$, \dots, $XXX$, $XXY$, $XYZ$.
\end{enumerate}
\end{lemma}
{\bf Proof.}
In view of applications in physics, we address two cases:
(1) Consider the scenario where we have full local controllability on each type of qubit and
a connected coupling topology of pairwise Ising (or Heisenberg) interactions (cp.~Tab.~I).
With $\fk'$ being trivial, the Lie closure of the controls {\em without} the
drift $\{i H_\nu\}\setminus i H_d$ is a {\em semi-simple} Lie algebra 
given in the irreducible representation
$\fh = \fh_1 \hoplus \fh_2 \hoplus\cdots\hoplus\fh_r \subsetneq \fk$,
where the sum typically extends over the local qubit spaces.
As the drift term shows a coupling topology represented by 
a connected graph, there is no subalgebra $\fh_k$
that remains invariant (in the sense of being normalised)
under all the coupling terms in $i H_d$. Thus the decomposition into a Kronecker
sum that exists for any semi-simple Lie algebra disintegrates when taking into account 
all the coupling terms and hence the Lie closure must turn into a {\em simple} Lie algebra 
at the latest upon including $i H_d$.

(2) Consider the scenario where we have full local control on a (non empty) subset of qubits while
others have no local control at all (cp.~Tab.~II). If one has again a connected coupling
topology such that all the uncontrolled qubits have at least one coupling of Heisenberg XXX 
(or XXY or XYZ) type, while the coupling between pairs of locally controlled qubits may
take Ising form as in (1), then observe the following: controlled local qubits 
transfer to their uncontrolled neighbours by the Heisenberg XXX (XXY, XYZ) interaction.
Note that a commutator of a single-qubit operator with $H_{xxx}$ acts like a \swap,
but the Heisenberg XXX interaction is entangling (cp the $\sqrt{\swap}$) and thus
does not allow for a Kronecker-sum structure giving a merely semi-simple system algebra;
it enforces simplicity.
Moreover if there is no symmetry in the coupling Hamiltonians, even several uncontrolled
qubits coupled to one single controlled qubit maintain irreducibility (by a trivial commutant
resulting from lack of symmetry) and simultaneously simplicity by coupling local 
qubit spaces (see Fig.~\ref{fig:tab2fig} (e) and Tab.~II). 
This generalises the scenario of \/`infecting\/' graphs \cite{Burg08}.
\hfill$\blacksquare$

\medskip
{\em Example:} Consider a two-qubit system with local controllability,
so the closure to $\{i H_\nu\}\setminus i H_d$ is $\su(2){\hoplus}\su(2)$, which is
semi-simple (and isomorphic to $\so(4)$) 
and has just a trivial commutant. Upon including, e.g., an Ising coupling,
the Lie closure of the full $\{i H_\nu\}$ turns into $\su(4)$, which is simple.

\medskip
So lack of symmetry gives a trivial commutant,
which in turn implies irreducibility and together with compactness 
it entails (at least) semi-simplicity, 
while a connected coupling topology on top finally ensures simplicity.

\begin{corollary}
Let a controlled closed $n$ spin-$\tfrac{1}{2}$ qubit quantum system be given which is 
characterised by the respective drift and control Hamiltonians 
$\{iH_\nu\} \subset \su(N)$ and their Lie closure
$\langle{i H_d, i H_j} \,|\,j=1,2,\dots,m\rangle_{\rm Lie} = \fk$
being a subalgebra to $\su(N)$, where $N\geq 4$. 

If a check on the $\{H_\nu\}$ ensures the only {\em candidates}
for a simple real compact Lie algebra $\fk\subseteq\su(N)$ are 
of the type $\mathfrak a_\ell$ $(\su(\ell+1))$ in $n$ spin-$\tfrac{1}{2}$
representation,
then the following assertions are equivalent:
\begin{enumerate}
\item[(a)] the system is fully controllable;
\item[(b)] the Lie closure is a representation $\fk \rep \su(N)$;
\item[(c)] the Lie closure $\fk$ is a simple Lie algebra and its commutant is trivial
	\begin{equation}
	\fk' = \{\lambda\cdot\unity_N\}\quad\text{with}\quad \lambda\in i\cdot\R{}\;;
	\end{equation}
\item[(d)] $\fk$ is an irreducible representation of a simple Lie algebra in
		$i\cdot \herm_0(N)$.
\end{enumerate}
\end{corollary}

{\bf Proof.}
For $(a) \Longleftrightarrow (b)$ see \cite{JS72}, while $(c) \Longleftrightarrow (d)$
is Schur's Lemma applied to the representation of Hilbert spaces, see, e.g., 
Proposition 2.3.8.~in \cite{BR1}.
The forward implication $(b) \Longrightarrow (c) \Longleftrightarrow (d)$ readily follows since
$ \su(N)$ is simple and the Lie closure 
$\fk = \su(N)$ leaves no non-trivial invariant subspace or
subalgebra of skew-Hermitian operators in $i\cdot\herm(N)$, hence it is {\em irreducible}. 

Only the backward implication $(c) \Longleftrightarrow (d) \Longrightarrow (b)$ is intricate
because $\su(N)$ is not unique among the 
simple compact Lie algebras: 
they comprise the classical types $\mathfrak a_\ell, \mathfrak b_\ell, \mathfrak c_\ell,
	\mathfrak d_\ell$ as well as the exceptional ones $\mathfrak e_\ell, \mathfrak f_\ell$
	and $\mathfrak g_\ell$ \cite{Helgason78},
all of which have to be ruled out along the lines of the subsequent {\em constructive candidate filter}
just leaving the type $\mathfrak a_\ell$.
\begin{lemma}[Candidate Filter]
Given a set of Hamiltonians $\{i H_\nu\}$ generating the Lie closure
$\fk \subseteq \su(2^n)$ with the promise that $\fk$ is an irreducible 
$n$ spin-$\tfrac{1}{2}$ representation 
of a compact simple Lie subalgebra to $\su(2^n)$ and $n\geq 2$.
Then $\fk = \su(2^n)$ unless one of the following three conditions is satisfied:
\begin{enumerate}
\item all the $\{i H_\nu\}$ are jointly conjugate to a set $\{i \widetilde H_\nu\}$,
	where each element is real and skew-symmetric $\widetilde H_\nu^t = - \widetilde H_\nu$, or
\item all the $\{i H_\nu\}$ are jointly conjugate to a set $\{i \widetilde H_\nu\}$,
	where each element is unitary and symplectic $J \widetilde H_\nu  = - \widetilde H_\nu^t J$
	with $J:=i\sigma_y\otimes \unity_{N/2}$ so $J^2=-\unity_N$, or
\item the $\{i H_\nu\}$ generate an irreducible unitary subalgebra $\su(N')\subsetneq\su(N)$
	with $N'<N$ that is compatible with an irreducible $n$ spin-$\tfrac{1}{2}$ representation.
\end{enumerate}
\end{lemma}

{\bf Proof of the Lemma.}
For multi-spin systems with $N\geq 4$ consider
the {\em compact} classical simple Lie algebras \cite{Helgason78}
as candidate subalgebras to $\su(N)$
\begin{eqnarray*}
\mathfrak a_\ell \; (\ell \geq 1) &:& \su(\ell + 1,\C{}) \\
\mathfrak b_\ell \; (\ell \geq 2) &:& \so(2\ell + 1,\R{}) \\
\mathfrak c_\ell \; (\ell    > 3) &:& \spp(\ell,\C{})\cap\uu(\ell,\C{})\quad\text{(with $\ell$ even)} \\
\mathfrak d_\ell \; (\ell \geq 4) &:& \so(2\ell,\R{}) \;.
\end{eqnarray*}
As $\ell\geq 3$, the isomorphisms 
(see, e.g., Thm.~X.3.12 in \cite{Helgason78}) 
$\mathfrak a_1 \sim \mathfrak b_1 \sim \mathfrak c_1$ 
and $\mathfrak b_2 \sim \mathfrak c_2$ or the isolated semi-simple case
$\mathfrak d_2 \sim \mathfrak a_1\hoplus \mathfrak a_1$
are of no concern. 
The odd-dimensional type $\mathfrak b_\ell$ is ruled out
as $\fk$ is an {\em irreducible simple} subalgebra to $\su(2^n)$ 
thus leaving the types $\mathfrak c_\ell$ and  $\mathfrak d_\ell$ as 
the only classical alternatives to $\mathfrak a_\ell$.
Note that option (3) is impossible by the promise of spin-$\tfrac{1}{2}$
representations: e.g., in $\su(4)$ there is an irreducible spin-$\tfrac{3}{2}$
representation of $\su(2)$ as a proper subalgebra ${\su(2)}_{j=3/2}\subsetneq\su(4)$,
but a change in spin quantum number $j$ and number of qubits $n$ 
is unphysical and ruled out by premiss.
So if $\fk$ is a classical proper simple subalgebra to $\su(2^n)$, then it
is conjugated to $\so(2^n)$ or to $\spp(2^{n-1})$,
in which case the $\{i H_\nu\}$ would be either jointly conjugate 
(1) to a set of skew-symmetric or (2) to a set of symplectic Hamiltonians 
as stated in the Lemma. 

The exceptional Lie algebras
$\mathfrak e_\ell, \mathfrak f_\ell, \mathfrak g_\ell$
can---by dimensionality---be excluded to count for {\em irreducible} 
$n$ spin-$\tfrac{1}{2}$ representations
of simple subalgebras to some $\su(2^n)$ in multi-qubit systems, as,
e.g., in the simplest representations \cite{Min06} (see also \cite{Baez01}) one has
\begin{equation*}
\fg_2\subset \so_7;\; \ff_4 \subset \so_{26};\; \fe_6\subset \sll_{27};\; \fe_7\subset\spp_{56};\; 
\fe_8\subset\so_{248}\quad.
\end{equation*}
This ends the (somewhat preliminary) proof of the Lemma and the Corollary.
\hfill$\blacksquare$

\begin{remark}
Finally suffice it to add to the above sketch
that---with the single exception of $\fg_2$ in an example of dimension $N$=7 
which again is of no concern in the qubit systems here---they 
fail to generate groups acting transitively 
on the sphere or on $\R N \setminus \{0\}$ as has been shown in \cite{DiHeGAMM08}
building upon more recent results in \cite{Kra03} to fill earlier work \cite{Bro73,BW79}.
\end{remark}

\bigskip
Fortunately, in practice the underlying simple Lie algebra may already be evident
from the context: consider for instance a typical spin coupling Hamiltonian
\begin{equation}
H_S = \sum A_{jk}(\sigma_{jx}\otimes\sigma_{kx}) + B_{jk}(\sigma_{jy}\otimes\sigma_{ky}) 
	+ C_{jk}(\sigma_{jz}\otimes\sigma_{kz})
\end{equation}
or a common quadratic Hamiltonian
\begin{equation}
H_{B(F)} = \sum\limits_{j,k=1}^n A_{jk} a_ja_k + B_{jk} a_ja_k^\dagger 
		+ C_{jk} a_j^\dagger a_k + D_{jk} a_j^\dagger a_k^\dagger\;,
\end{equation}
where the $a_j$ and $a_j^\dagger$ denote the respective annihilation and creation operators
satisfying the canonical (anti)commutation relations
\begin{eqnarray}
\text{bosons:\;} \protect{[a_j,a_k^\dagger]} &= \delta_{j,k} \quad\text{and}\quad \protect{[a_j,a_k]} &= 0\\
\text{fermions:\;} \protect{\{a_j,a_k^\dagger\}} &= \delta_{j,k} \quad\text{and}\quad \protect{\{a_j,a_k\}} &= 0\;.
\end{eqnarray}
Therefore  we get the natural representations
\begin{alignat}{6}
&\text{spin systems}& :  &\quad\mathfrak a_\ell\quad& : &\quad\{iH_\nu\}\in \su(\ell+1)&\\
&\text{bosonic systems}& :  &\quad\mathfrak c_\ell\quad& : &\quad\{iH_\nu\}\in \spp(\ell)&\\
&\text{fermionic systems}& :  &\quad\mathfrak d_\ell\quad& : &\quad\{iH_\nu\}\in \so(2\ell)&\quad.
\end{alignat}

\noindent
Thus the ambiguity that arises when given $\fk$ with the only promise it is an
irreducible unitary representation of some simple compact Lie algebra $\fk\subseteq\su(2^n)$
has its origin in physics:
the unitary representation as a (pseudo)spin system could simulate 
or encode by conjugation 
\begin{enumerate}
\item a true spin system, 
\item a bosonic system, or
\item a fermionic system. 
\end{enumerate}
This ambiguity is removed as soon  as the (pseudo)spin Hamiltonians $\{i H_\nu\}$ can be 
checked for either {\em allowing} or for safely {\em excluding} a bosonic or a fermionic 
representation.
If not, a purely numerical assessment is more tedious.

\noindent
{\em Check for Symplectic Representation~\cite{BW79} :}
Determine all solutions $\mathcal S:=\{J_i\}$ to the coupled system 
$J^t + J =0$ with $JH_\nu + H^t_\nu J = 0$ for all $\{H_\nu\}$.
If $\mathcal S=\{0\}$ there is no equivalent symplectic representation
of $\{iH_\nu\}\subseteq \fk$. However, if there exists a non-zero element in $\mathcal S$
that is non-singular, then $\fk$ is a subalgebra to $\su(N)$ that is conjugate to $\spp(N/2)$.

\noindent
{\em Check for Skew-Symmetric Representation:} 
Make use of the $vec$-notation \cite{HJ2}, where the product of matrices
$AXB$ becomes $(B^t\otimes A) \vec(X)$ with $\vec(X)$ being the
column vector of all columns in the matrix $X$. Check
whether there is a joint unitary matrix $V\in U(N)$ such that 
$(V^tV\otimes V^\dagger \bar V) \vec(H^t_\nu) = - \vec(H_\nu)$ for all $\nu$.


\subsubsection*{Example 3: Lack of Symmetry}

\begin{figure}[Ht!]
\begin{centering}
\includegraphics[scale=0.5]{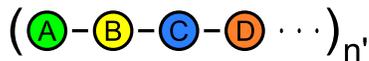}
\par\end{centering}
\caption{Example of a periodic spin chain of Heisenberg or Ising coupling type 
in which the spin-type pattern of type-wise jointly controllable spins 
is repeated $n'$ times.}
\label{fig:spinchain-periodic}
\end{figure}

Consider spin chains of Ising or Heisenberg coupling type 
with repeated patterns of spin types than can only be controlled 
typewise. Let the patterns be arranged 
such that no $S_2$ mirror symmetry occurs. An example is 
illustrated in Fig.~\ref{fig:spinchain-periodic}. Then 
all instances of non-symmetric spin chains (and other connected coupling topologies)
numerically addressed thus far have clearly reproduced $\fk=\su(2^n)$
by {\bf Algorithm 1}
once an  independent algebraic analysis had ensured that
the commutant to the generating set of drift and control Hamiltonians $\{H_\nu\}'$ 
is trivial.

%

\section{Observability}
Controllable systems are closely related to \/`observable\/' ones, as has been
known for a long time in classical systems and control theory, see, e.g., \cite{Sontag}.
In the context of quantum dynamics, the interrelation has been summarised in
the textbook by D'Alessandro~\cite{dAless08}. There the notion of observability
is in a direct sense.
Consider an observed bilinear control system (henceforth termed $\Sigma$)
\begin{equation}\label{eqn:bilobs1}
\dot{\rho}(t) = -i [H_u,\rho(t)]\quad\text{where}\quad \rho(0)=\rho_0
\end{equation}
with the measured output by the detection observable $C$
\begin{equation}\label{eqn:bilobs2}
y(t):= \tr\{C^\dagger \rho(t)\}\quad.
\end{equation}
Recall that we specified the reachable set in terms of subgroup orbits 
generated by the dynamic Lie algebra, so
\begin{equation}
\rho(t)\in \mathcal O_\bK(\rho_0) =\{K\rho_{0}K^{-1}|\, K\in \bK = \exp \fk\}\quad.
\end{equation}
Then two states represented as density operators $\rho_{A0}$ and $\rho_{B0}$ are
said to be {\em indistinguishable} by the observable $C$, if at any time $t$ and under any control $u(t)$
they give the same expectation values $\langle C\rangle (t)$ over the entire
reachable sets, i.e.~if
\begin{equation}
\tr\{C^\dagger \mathcal O_\bK(\rho_{A0})\} = \tr\{C^\dagger \mathcal O_\bK(\rho_{B0})\}\quad.
\end{equation}
Thus the notion of indistinguishability is closely related to the topic of
the {\em $C$-numerical range} \cite{GS-77,Li94} being defined as
\begin{equation}
W(C,A) := \tr \big\{(C^\dagger UAU^\dagger)\; |\; U\in SU(N)\big\}\quad.
\end{equation}
In particular,
the set of expectation values $\langle C\rangle (t)$ exactly coincides with
what we introduced as the {\em relative $C$-numerical range} \cite{WONRA_tosh,WONRA_dirr}
\begin{equation}
W_{\mathbf K}(C,A) := \tr \big\{(C^\dagger KAK^\dagger)\; |\; K\in \mathbf K \subsetneq SU(N)\big\}
\end{equation}
with respect to the compact connected subgroup $\mathbf K$ generated by the dynamic 
Lie algebra $\fk$ of the controlled system. Therefore one obtains the following:

\begin{corollary}
Two states $\rho_{A0}$ and $\rho_{B0}$ are indistinguishable by an observable $C$
if their relative $C$-numerical ranges with respect to the (compact connected) dynamic Lie group 
$K:=\exp\fk\subseteq SU(N)$ coincide
\begin{equation}
W_{\mathbf K}(C,\rho_{A0}) = W_{\mathbf K}(C,\rho_{B0})\quad.
\end{equation}
\end{corollary}

\medskip
Now a controlled quantum system is called {\em observable} whenever two states 
are indistinguishable if and only if they are equal. In Ref.~\cite{dAless08}
D'Alessandro showed that fully controllable {\em quantum} systems are also 
observable ({\em ibid.}, Proposition 4.2.8). This is the quantum adaptation 
of its famous and even stronger 
forefather in classical systems and control theory, where controllability
and observability have long been established to be in a duality relation:
i.e.~a classical system is controllable if and only if it is observable.
For details see, e.g., \cite{Sontag}.

\subsection{Constraints by Symmetry}

Here we resume some of the arguments of \cite{dAle03, dAless08} and relate them 
to symmetries of quantum systems:
to this end turn the observables $C\in \herm(N)$ into their traceless forms 
$\Tilde C \in \herm_0(N)$
\begin{equation}
\Tilde C:= C - \tfrac{\tr C}{N}\cdot\unity_N\quad.
\end{equation}
Then each observable comes with an {\em observability space} 
with respect to the dynamic Lie algebra $\fk$, which in analogy to \cite{dAle03} 
can be written as
\begin{equation}
\mathcal V_C := \bigoplus\limits_{j=0}^\infty \adr_\fk^j {\rm span}\{i\tilde C\}\quad,
\end{equation}
where $\adr_\fk^j$ operating on $C$ is defined by the linear span of $j$ times repeated Lie brackets
of elements $k_i\in\fk$ with $iC$. Due to the Cayley-Hamilton theorem, the infinite
sum terminates at some finite value. $\mathcal V_C$ is a subspace of $\su(N)$,
but not necessarily a subalgebra.
The observability space (and its orthocomplement $\mathcal V^\perp_C$ of $\mathcal V_C$ in $\mathfrak u(N)$)
both form ideals under the dynamic Lie algebra
\begin{equation}
[\fk, \mathcal V_C] \subseteq \mathcal V_C\quad\text{and}\quad
		[\fk, \mathcal V^\perp_C] \subseteq \mathcal V^\perp_C\quad.
\end{equation}
Moreover in ref.~\cite{dAless08} it was also shown that two states $\rho_{A0}$ and $\rho_{B0}$
are indistiguishable by $C$ if and only if their difference lies in the
orthocomplement $(\rho_{A0} - \rho_{B0})\in i \mathcal V^\perp_C$.
This appears very natural, as the difference between distinguishable
states has to be observable.
So the controlled system  $\Sigma$ of Eqn.~\eqref{eqn:bilobs1} with 
output as in Eqn.~\eqref{eqn:bilobs2} is observable
by $C$  if 
\begin{equation}
\mathcal V_C \rep \mathfrak{su}(N)\quad.
\end{equation}

\medskip
By similar arguments, one may extend the controlled bilinear system to be
observed by several non-trivial detection operators 
$\mathcal C := \{C_1,C_2,\dots,C_r\}$, see \cite{dAle03}
for details. Since the respective observability space $\mathcal V_{\mathcal C}$ is 
again an ideal under the dynamic Lie algebra $\fk$, and full controllability
entails $\fk = \su(N)$, then $\fk$ is a
simple Lie algebra by construction and has only trivial ideals,
$\su(N)$ itself and the identity---which correspond to
$\mathcal V_\mathcal C$ and $\mathcal V^\perp_\mathcal C$, 
respectively.
Therefore a fully controllable system is also observable \cite{dAless08}.

Recall that the commutant $\fk'$ to the dynamic Lie algebra $\fk$
constitutes an ideal to $\fk$.
As in Sec.~I, non-trivial symmetries preclude that the dynamic Lie algebra is simple.
An immediate consequence is that non-trivial symmetries $s\in\fk'$
not only exclude full controllability but also restrict observability,
because the orthocomplement $V^\perp_\mathcal C$ becomes non trivial.
This can be made more precise by the following natural definition:
\noindent
Let $\Sigma$ be again a bilinear quantum control system with output
observed by a given set of detection operators $\mathcal C:=\{C_1,C_2,\dots,C_r\}$
that shall be mutually orthogonal with respect to the Hilbert-Schmidt scalar product.
Let $\mathcal V(\Sigma,\fk, \mathcal C)$ 
be the associated observability space under the dynamics of $\Sigma$
governed by the Lie algebra $\fk$.
Then the system is called {\em observable by $\mathcal C$ on the subspace} $\mathcal R\subseteq \herm(N)$
if $\mathcal R:= {\rm span}_{\R{}}\{\rho_1,\rho_2,\dots,\rho_\ell\}$,
where $\{\rho_1,\rho_2,\dots,\rho_\ell\}$ is a {\em maximal set} of linearly
independent states
such that for all pairs $(\rho_\nu,  \rho_\mu)\in \mathcal R \times \mathcal R$
\begin{equation}
(\rho_\nu - \rho_\mu) \notin i\;\mathcal V^\perp(\Sigma,\fk, \mathcal C)\quad.
\end{equation}
Clearly, the observable subspace $\mathcal R$ grows with the 
number of (experimentally) available detection operators $\mathcal C$,
or more precisely, with the number of disjoint orbits under the dynamic group $\bK$,
i.e. $\mathcal O_\bK(C_\nu)$ with $C_\nu \in\mathcal C$.
Now if the observables $\mathcal C$ share the same symmetry group as the dynamic system
and hence its algebra $\fk$, then the observability space must be a subspace to $\fk$.

\medskip
\noindent
Therefore in actual experiments we have the following
\begin{proposition}
Let $\Sigma$ be a bilinear quantum control system with 
associated dynamic Lie algebra $\fk$.
Then an operator $A\in\herm(N)$ is observable by the
set of detection operators $\mathcal C$ if to the
associ\-ated observability space $\mathcal V(\Sigma,\fk, \mathcal C)$
there is a set of preparable linearly independent initial states of the system 
$\{\rho_1,\rho_2,\dots,\rho_\ell\}$ spanning $\mathcal R$
so that 
(i) $\exists A(t)\in \mathcal O_\bK(A): A(t)\in\mathcal R$ and
(ii) $\Sigma$ is observable by  $\mathcal C$ on $\mathcal R$.
\end{proposition}

The consequences for quantum system identification are immediate and, fortunately,
not limited by the intrinsic symmetry of the system:
\begin{corollary}
A closed controlled quantum system $\Sigma$ with dynamic Lie algebra $\fk$
can be identified by a set of orthogonal detection operators $\mathcal C:=\{C_1,C_2,\dots,C_r\}$ 
in the following two scenarios:
\begin{enumerate}
\item if the system Hamiltonian $H_0$ is entirely unknown, then it is sufficient that
	the observability space associated to the observables $\mathcal C$ is
	$\mathcal V(\Sigma,\fk, \mathcal C) = \mathfrak{su}(N)$;
\item if only the structure of the system Hamiltonian $H_d$ is known in the sense
	it is a linear combination of orthogonal basis elements, where the knowledge about
	the expansion coefficients is confined to \/`zero\/' or \/`non-zero\/',
	then it suffices all the basis elements with non-zero coefficients
	are observable by $\mathcal C$.
\end{enumerate}
\end{corollary}
\noindent
In a sense, scenario (1) mirrors full controllability, whereas scenario (2)
reflects \/`task controllability\/' with respect to all non-zero Hamiltonian components. ---
In relation to the second scenario note that the dynamic Lie algebra $\fk$ is itself 
independent of the actual coupling strengths as long as they are sizeably different from zero
(and do not decide whether a symmetry is introduced or is broken).


Up to now, we have treated observability in one step. 
However, similar arguments generalise
naturally to several measurements at $k$ different times \cite{dAle03} and 
thus form a basis of quantum process tomography. For instance, then the notion of 
being indistinguishable is replaced by being indistinguishable in $k$ time steps.

Moreover it is to be anticipated that repeated measurements will play a further
role in indirect Hamiltonian identification as introduced recently \cite{Burg09}
to determine the parameters of a system Hamiltonian by tracing time evolutions 
measured on just a subset of gateway qubits.

\subsection{Constraints by Relaxation}
Consider a controlled bilinear system $\Sigma_M$ with Markovian relaxation
expressed in GKS-Lindblad form 
\begin{equation}\label{Eq:CLKE}
\dot \rho = -\ri \adr_{H_u}(\rho) - \Gamma_{L}(\rho)\quad,
\end{equation}
where the output shall be measured by the observable $C$
\begin{equation}\label{eqn:CLKE-obs}
y(t) = \tr\{C^\dagger \rho(t)\}\quad.
\end{equation}
Controllability and observability issues in those type of 
controlled open quantum system
are more intricate as is evident even in the simplest case: 
Let the $\ri \adr_{H_j}$ be skew-Hermitian, while $\Gamma_L$ shall 
simplify to be Hermitian (which generically need not hold).
The usual Cartan decomposition of
$\mathfrak{gl}(N^2,\C{}) := \mathfrak k \oplus \mathfrak p$
into skew-Hermitian matrices ($\mathfrak k$) and Hermitian matrices
($\mathfrak p$) with its commutator relations
$[\mathfrak k, \mathfrak k]\subseteq \mathfrak k, \;
[\mathfrak p, \mathfrak p]\subseteq \mathfrak k, \;
[\mathfrak k, \mathfrak p]\subseteq \mathfrak p$
illustrates that double commutators of the form
\begin{equation}
\big[[\ri\adr_{H_j}, \Gamma_L], [\ri \adr_{H_k}, \Gamma_L]\big]
\end{equation}
may generate \emph{new} $\mathfrak k$-directions in the dynamic Lie algebra that
by relaxation turns into a subalgebra $\fs\subseteq \gl(N^2,\C{})$.

As we have recently illustrated on a general scale \cite{DHKS08},
controlled open systems thus fail to comply with
the standard notions of controllability, see also
\cite{DiHeGAMM08, Alt03}. 
If, in the absence of relaxation, the Hamiltonian system
is fully controllable, one finds
\begin{equation}
\label{eqn:fullcontrol}
\langle \ri H_d,\ri H_j\;|\; j=1,\dots,m\rangle_{\sf Lie} =
\mathfrak{su}(N) \;,
\end{equation}
while master equations like Eqn.~\eqref{Eq:CLKE}
generically give a dynamic Lie algebra
\begin{equation}
\begin{split}
\mathfrak s_{\rm open} &:=\langle  \ri\ad{H_d} + \Gamma_L, \ri\ad{H_j}
		\;|\; j=1,2,\dots,m \rangle_{\sf Lie} \\[2mm]
	&\phantom{:}= \gl(\mathfrak{her}_0(N))\;,
\end{split}
\end{equation}
cf.~\cite{DHKS08,DiHeGAMM08,Alt04,Diss-Indra}.
Then the dynamics of the entire open system takes the form of a
contraction semigroup contained in $GL(\mathfrak{her}_0(N))$;
the relaxative part interferes with the coherent Hamiltonian part
generating new directions in the Lie algebra, where the geometry
of the interplay determines the set of explored states.

Hence we introduced weaker controllability concepts in open systems \cite{DHKS08}:
fix the subalgebras generated by the control terms as
\begin{equation}
\mathfrak k_{c}
:= \expt{\ri H_{1}, \dots, \ri H_{m}}_{\rm Lie}
\end{equation}
and the one extended by the  \emph{Hamiltionian drift} as
\begin{equation}
\mathfrak k_{d}
:= \expt{\ri H_{d}, \ri H_{1}, \dots, \ri H_{m}}_{\rm Lie}\quad.
\end{equation}
Then an open Markovian quantum system $\Sigma_M$ \eqref{Eq:CLKE} is
\begin{enumerate}
\item[(a)]
{\em Hamiltonian-controllable} ({\sc h}), if $\mathfrak k_c = \su(N)$ and
no bounds on the control amplitutes \mbox{$u_j$, $j = 1, \dots, m$} are imposed 
(idealised instantanous reachability);
\item[(b)]
{\em weakly Hamiltonian-controllable} ({\sc wh}), provided
$\mathfrak k_{d} = \su(N)$ and
$\Gamma_L\big|_{\mathfrak{her}_0(N)} \!=\! \gamma \unity$
with $\gamma \geq 0$.
\end{enumerate}
In general, an open quantum system that is fully controllable
in the absence of relaxation need not be
{\sc wh}-controllable when including relaxation. 

\noindent
Now with regard to observability we obtain
\begin{corollary}
Let $\Sigma_M$ be a bilinear controlled open Markovian quantum system following 
Eqns.~\eqref{Eq:CLKE} and \eqref{eqn:CLKE-obs}. Assume that in the
absence of relaxation, the system $\Sigma$ is fully controllable and thus
comes with the associated observability space $\mathcal V(\Sigma,\fk,C)=\su(N)$. 
Then
\begin{enumerate}
\item if it is {\sc h}-controllable, the open system has the same observability space 
$\mathcal V(\Sigma_M,\fk,C)=\mathcal V(\Sigma,\fk,C)$;
\item if it is {\sc wh}-controllable, the open system has an observability space that is the
contraction of $\su(N)$ in $\gl(N)$;
\item generically, the open system has an observability space that is 
a subset in $\gl(\mathfrak{her}_0(N))$ growing with time after which the
measurements are taken.
\end{enumerate}
\end{corollary}
\noindent
We recently gave an illustration of the corresponding reachable set in 
generic open Markovian systems \cite{DHKS08} showing how intricate it is
to specify reachable sets in generic open systems. Hence addressing observability
spaces in open systems is an equally subtle enterprise with lots of open
research problems.

\bigskip

\section{Conclusions and Outlook}
We have treated controllability and observability in a unified Lie-algebraic framework
incorporating constraints by symmetry for closed systems and constraints by relaxation
in open ones. In particular, the dynamic system Lie algebra allows for specifying the reachability sets
of closed systems explicitly. In symmetric quantum hardware setups the feasible tasks in quantum simulation
or quantum gate synthesis can be made precise thus giving valuable guidelines for quantum
hardware design matched to solve given problems.

Likewise, the dynamic system Lie algebra determines those Hilbert subspaces
that can be directly observed with a given set of detection operators. We anticipate
that the approach will prove widely useful in Hamiltonian system identification as well 
as in process tomography.

Using the Lie cone as the open-system generalisation of the dynamic system Lie algebra
in closed ones seems a logical consequence, yet in practice it is by no means straightforward.
Thus in \cite{DHKS08} we gave a sketch how to approach it from the inside as well as from the outside.

\medskip
\begin{acknowledgments}
This work was supported in part by the integrated EU programme QAP,
by the International Doctorate Program of Excellence
{\em Quantum Computing, Control, and Communication} (QCCC)
by the Bavarian excellence network ENB, and by
{\em Deutsche Forschungsgemeinschaft} (DFG) in the
collaborative research centre SFB 631.
We wish to thank Daniel Burgarth for fruitful discussion in particular
on the bicommutant. We are also indebted to Gunther Dirr and Robert Zeier for
stimulating exchange. Andreas Sp{\"o}rl was helpful in numerical
issues at the initial stage of the project.
\end{acknowledgments}

\subsubsection*{Note Added in Proof} 
After the first version of this paper was posted on
the preprint {\sf arXiv}, an article by the group of Tannor \cite{PST09} (with no preceeding preprint)
was published. It classifies all {\em non}-controllable multi-level systems by their Lie algebras in
terms of branching rules.
Here in Sec.~I-D we took the opposite approach and showed under which mild conditions one may
exclude other proper irreducible subalgebras in systems with no symmetry in order to obtain {\em full}
controllability by symmetry arguments rather than by the computationally costly Lie-algebra rank condition.

\bibliography{control21}

\end{document}